\documentclass[preprints,article,accept,moreauthors,pdftex]{Definitions/mdpi-arXiv}

\newcommand{\declarecommand}[1]{\providecommand{#1}{}\renewcommand{#1}}
\declarecommand{\q}{\mathbf{q}}
\declarecommand{\vk}{\mathbf{k}}
\declarecommand{\p}{\mathbf{p}}
\declarecommand{\Q}{\mathbf{Q}}
\declarecommand{\eps}{\varepsilon}
\declarecommand{\su}{\uparrow}
\declarecommand{\sd}{\downarrow}
\declarecommand{\sgn}[1] {\mathrm{sgn}\left({#1}\right)}
\declarecommand{\sgnnobr}[1] {\mathrm{sgn}{#1}}
\declarecommand{\sign}{\mathrm{sgn}}
\declarecommand{\abs}[1] {\left|{#1}\right|}
\declarecommand{\ii}{{\mathrm{i}}}

\firstpage{1}
\pubvolume{15}
\issuenum{15}
\articlenumber{0}
\pubyear{2022}
\copyrightyear{2022}
\externaleditor{{Academic Editors:} Liwei Geng and Yongke Yan}
\datereceived{29 July 2022}
\dateaccepted{1 September 2022}
\datepublished{}
\hreflink{https://doi.org/} 

\Title{Direct Observation of the Spin Exciton in Andreev Spectroscopy of Iron-Based Superconductors}

\TitleCitation{Direct Observation of the Spin Exciton in Andreev Spectroscopy of Iron-Based Superconductors}

\Author{Maxim M. Korshunov 
$^{1,}$*\orcidA{}, Svetoslav A. Kuzmichev $^{2,3}$\orcidB{} and Tatiana E. Kuzmicheva $^{3}$\orcidC{}} 

\AuthorNames{M.M. Korshunov, S.A. Kuzmichev, and T.E. Kuzmicheva}

\AuthorCitation{Korshunov, M.M.; Kuzmichev, S.A.; Kuzmicheva, T.E.}

\address{%
$^{1}$ \quad Kirensky Institute of Physics, Federal Research Center KSC SB RAS, Akademgorodok, \linebreak 660036 Krasnoyarsk, Russia\\ 
$^{2}$ \quad Faculty of Physics, Lomonosov Moscow State University, 119991 Moscow, Russia; kuzmichev@mig.phys.msu.ru\\
$^{3}$ \quad Lebedev Physical Institute, Russian Academy of Sciences, 119991 Moscow, Russia; kuzmichevate@lebedev.ru}

\corres{Correspondence: mkor@iph.krasn.ru 
.
}

\abstract{Quasiparticle excitations provide viable information on the physics of unconventional superconductors. Higgs and Leggett modes are some of the {classic} examples. Another important bosonic excitation is the spin exciton originating from the sign-changing superconducting gap structure. Here we report a direct observation of the temperature-dependent spin exciton in the Andreev spectra of iron-based superconductors. Combined with the other experimental {evidence}, our observation confirms the extended $s$-wave ($s_\pm$)
order parameter symmetry and indirectly proves the spin-fluctuation mechanism of Cooper pairing.}



\begin{document}

\section{Introduction}

Unconventional spin-singlet superconductivity is commonly characterized by a sign-changing gap. It naturally has a nodal structure, either lines of zeros or point nodes. Depending on whether the nodal lines of the gap cross the Fermi surface, the {thermodynamic} properties change. Normal quasiparticles located within the nodal structure contribute to the low-temperature specific heat, London penetration depth, and thermal conductivity. Those can be measured and the unconventional nature of the gap can be established. There are, however, cases with the nodeless sign-changing gap structure. The obvious 
example is the $s_\pm$ gap scenario for iron-based materials~\cite{Mazin2008}. The order parameter there has one sign in the center of the two-dimensional Brillouin zone and the opposite sign near the edges. {The} Fermi surface has multiple sheets and some of them are located near the center while others are near edges. They do not cross the order parameter's lines of zeros, thus the Fermi surface is fully gapped. At the same time, the gap on the central Fermi surface sheets (hole pockets) and the gaps on the other sheets (electron pockets) have opposite signs. To detect such a state, one has to perform some phase-sensitive experiment. The analogy of the seminal SQUID test~\cite{Tsuei1997} for the $d_{x^2-y^2}$ gap symmetry in cuprates is not applicable to the $s_\pm$ state in Fe-based superconductors (FeBS) because the state belongs to the $C_4$-symmetric $A_{1g}$ representation and is not sensitive to the 90$^\circ$ mutual orientation of two samples. Another approach is connected to the momentum-dependent structure of the dynamical spin susceptibility $\chi(\mathbf{q},\omega)$. Since the $s_\pm$ gap changes sign at some specific momenta $\mathbf{Q}$, 
spin susceptibility $\chi(\mathbf{q}=\mathbf{Q},\omega)$ diverges and {produces} the so-called spin resonance \linebreak peak~\cite{KorshunovEreminResonance2008,Maier2008}. Such a particle-hole bosonic excitation inside the spin gap of the superconducting state---a spin exciton---can be observed in the inelastic neutron scattering, and the spin resonance peak was indeed found in many iron pnictides and chalcogenides~\cite{LumsdenReview,Dai2015,Inosov2016}. To independently prove that the observed feature is the spin exciton, one has to consider an alternative way to probe the bosonic excitations in FeBS. Such an opportunity comes from the Andreev scattering where the Copper pair {breaks} into the particle and the hole. The measured conductance is affected by the scattering on bosonic excitations, thus, the latter can be detected.

Here we combine experimentally measured incoherent multiple Andreev reflection effect (IMARE) data on GdO$_{0.88}$F$_{0.12}$FeAs 
and theoretical calculations to show that the specific anomalous contributions to the measured conductance is directly related to the spin exciton thus confirming the sign-changing gap structure in FeBS.

\section{Materials and Methods}

The studied GdO$_{0.88}$F$_{0.12}$FeAs polycrystalline samples (hereafter Gd-1111) with almost optimal composition and critical temperatures $T_c \approx 49$\,K were prepared under high pressure. The details of synthesis and characterization of the samples are presented in Ref.~\cite{Khlybov2009}.

IMARE occurs in a ballistic SnS (superconductor--thin normal metal--superconductor) junction~\cite{Octavio1983,Averin1995,Kummel1990,Gunsenheimer1994}. For a ``long'' high-transparent SnS junction with $\xi < d < l$ (where $\xi$ is the superconducting coherence length, $d$ is the dimension of the metallic constriction, $l$ is {electron} scattering length)
below $T_c$ incoherent Andreev transport causes an excess current at any $eV$ which drastically rises at low bias voltages (so called foot feature) and a series of dynamic conductance dips called subharmonic gap structure (SGS). At any temperature $T$, the position of SGS is directly related to the gap magnitude $\Delta(T)$~\cite{Octavio1983,Kummel1990,Gunsenheimer1994} as: 

\vspace{-6pt}
\begin{equation}
eV_n(T) = \frac{2\Delta(T)}{n}, 
\end{equation}
\vspace{-6pt}

\noindent where $n=1,~2,\dots$ is the natural number representing the subharmonic order. In order to make mechanically controlled planar SnS junctions for {the} Andreev spectroscopy experiment, we used a break-junction technique~\cite{Moreland1985,Kuzmichev2016Review}. The method implies a cleavage of a layered sample along the crystallographic $ab$-planes at low temperatures, with the current flowing along the $c$-direction through the resulting planar break junction~\cite{Kuzmichev2016Review}.

\section{Results and Discussion}

During MARE observed here, in general, an electron could {lose} or gain its energy by coupling to a bosonic mode. At low temperatures, boson emission is likely and the bosonic mode energy $\varepsilon_0$ should be below $2\Delta(0)$
to be observable. A resonant interaction with a characteristic bosonic mode with the energy $\varepsilon_0$ causes a fine structure in the dI(V)/dV spectrum. Accompanying each Andreev dip, at higher bias, less-intensive satellite dip appears at position:

\vspace{-6pt}
\begin{equation}
eV_n = \frac{2\Delta + \varepsilon_0}{n},
\end{equation}
\vspace{-6pt}

\noindent forming an additional subharmonic series. The resulting fine structure looks similar to the case of microwave irradiated SnS junction observed in YBaCuO~\cite{Zimmermann1996}.

\begin{figure}[H]
\includegraphics[width=0.8\linewidth]{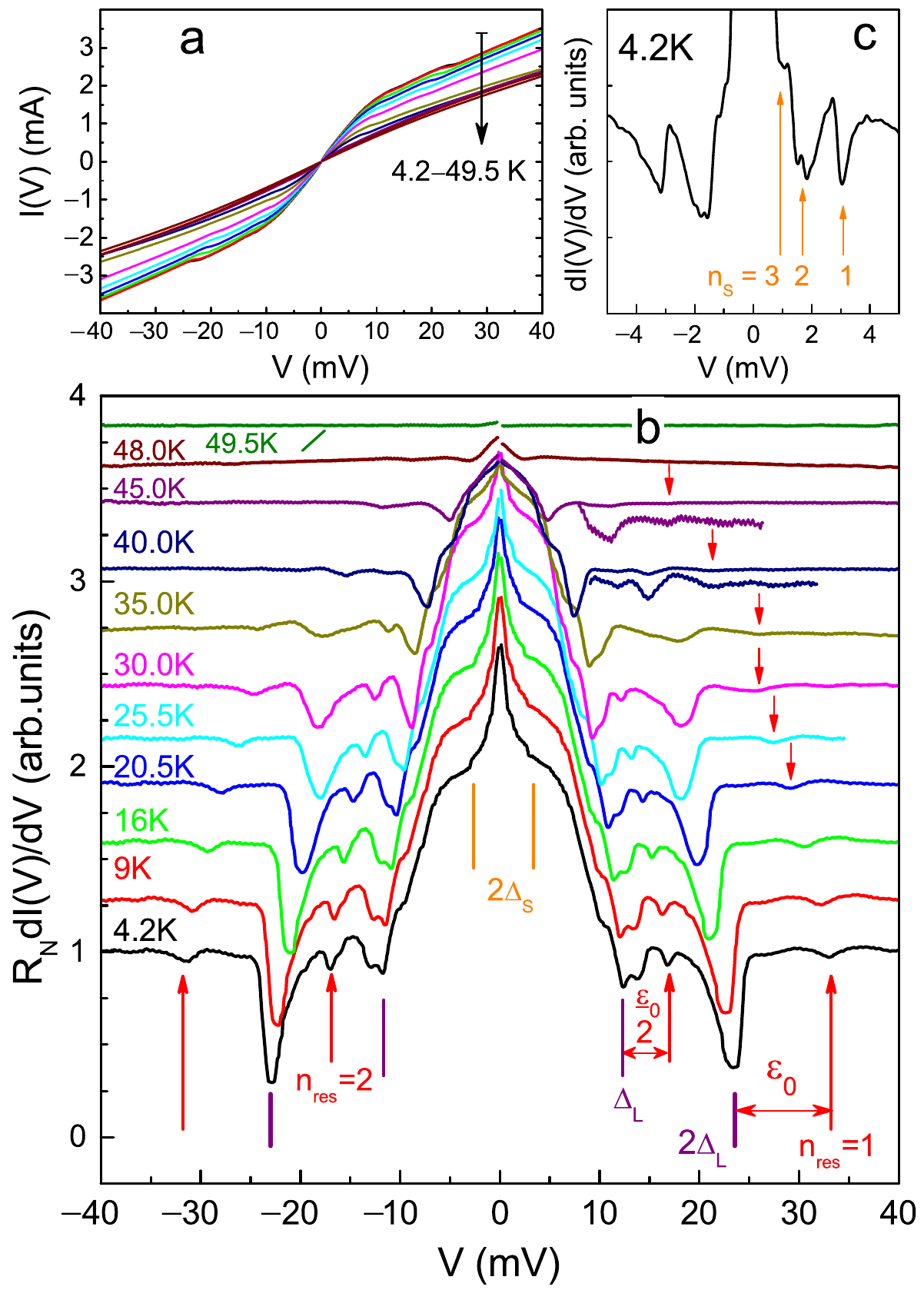}
\caption{Temperature 
evolution from $4.2$ K till the local critical temperature $T_c^\mathrm{local} \approx 49$\,K of current-voltage {characteristic}
({\bf a}) and dynamic conductance spectrum ({\bf b}) of SnS Andreev junction formed in Gd-1111. Andreev dips for the large  superconducting gap $\Delta_L(0) \approx 11.7$ meV are pointed to by purple vertical lines, satellite bosonic resonances are marked by red arrows and labels $n_\mathrm{res} = 1,~2$, and double arrows indicate the value of the boson energy $\varepsilon_0 \approx 10.4$ meV at 4.2 K. At 40 K and 45\,K, {vertically zoomed} fragments of dI(V)/dV detailing the boson resonances are also shown. dI(V)/dV curves are shifted vertically for clarity. $dI(eV > 2\Delta_L)/dV \to G_N = const$. Monotonic background was suppressed. ({\bf c}) The low-bias fragment of the dI(V)/dV spectrum at 4.2 K with additional background suppression that details the SGS of the small gap $\Delta_S(0) \approx 1.6$\,meV ($n_S = 1,~2,~3$ labels). \label{temp}}
\end{figure}

Current-voltage characteristics (CVC) measured at various temperatures are shown in Figure~\ref{temp}a. As compared with the CVC measured at $T \approx 49.5$\,K that is above $T_c$, the I(V) curves in the superconducting state show (i) a pronounced excess current in the whole $eV$ range, which drastically rises close to the zero bias (foot) and (ii) no supercurrent branch thus indicating a high-transparency (80\%--95\%) IMARE regime~\cite{Octavio1983,Averin1995}. Following the simple estimation presented in Ref.~\cite{Kuzmicheva2013} and using the normal resistance of the junction $R_N \approx 17$\,Ohm, one obtains the ratio of the mean free path to the contact dimension $l^{el}/d \approx 2.5$. Therefore, one should expect ballistic transport through the junction and observation of 2--3 Andreev subharmonics of each superconducting gap~\cite{Gunsenheimer1994}.

Temperature evolution of the corresponding dynamic conductance is shown in \linebreak Figure \ref{temp}b. Note that the spectra for different temperatures are shifted vertically for clarity. In reality, their conductance $dI(eV > 2\Delta_L)/dV$ tends to the normal-state $G_N$, which is nearly constant with the variation of temperature that also favors the ballistic regime. The spectrum measured at 49.5\,K becomes flat, thus determining the local critical temperature of the junction $T_c^\mathrm{local} \approx 49$\,K that corresponds to the transition to the normal state of the contact area.

\textls[20]{In the spectrum measured at 4.2\,K, the position of clear dips located at \linebreak $|eV| \approx 23.2$\,meV and $|eV| \approx 11.8$\,meV (purple vertical dashes in Figure \ref{temp}b) determine the large gap $\Delta_L(0) \approx 11.7$\,meV. At lower bias, the second SGS corresponding to the small gap is present (vertical orange bars). In order to detail it, we show the low-bias fragment of the dI(V)/dV-spectrum measured at 4.2\,K with additional monotonic background suppression in panel (c). The minima located at $|eV| \approx 3.2$\,meV, 1.6\,meV, and 1.07\,meV are interpreted as $n_S = 1,~2,~3$ subharmonics of the small gap $\Delta_S(0) \approx 1.6$\,meV. The resulting characteristic ratio for the large gap $2\Delta_L(0)/k_BT_c \approx 5.7$ exceeds the weak-coupling limit $3.5$, whereas the ratio for the small gap $2\Delta_L(0)/k_BT_c \approx 0.75$ appears well below $3.5$, which is typical for a ``weak'' condensate in a multiple-band superconductor.}

\begin{figure}[H]
\includegraphics[width=0.8\linewidth]{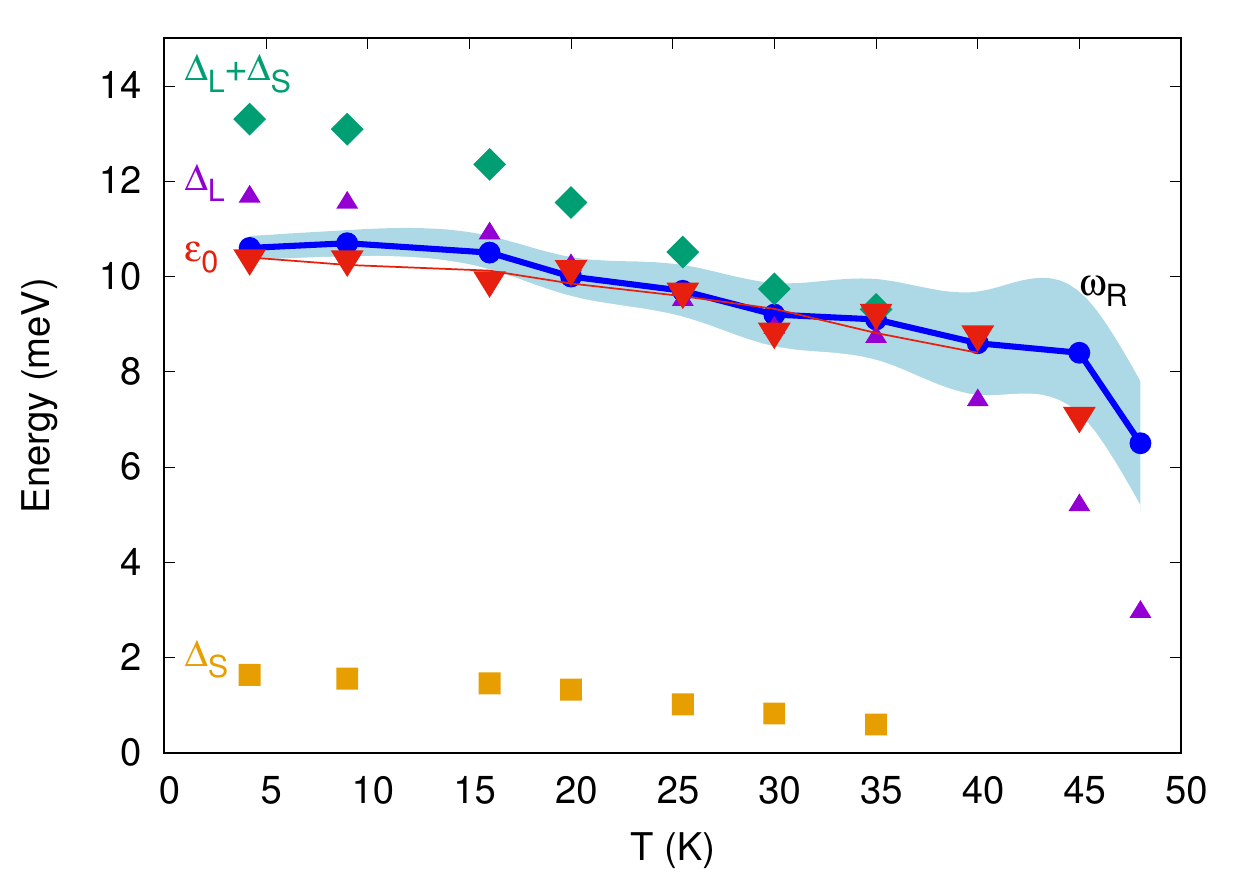}
\caption{Temperature dependence of the measured large gap $\Delta_L$ ({violet triangles}), measured and extrapolated small gap $\Delta_S$ ({orange squares}), their sum $\Delta_L+\Delta_S$ (green rhombuses), and the measured bosonic energy $\varepsilon_0$ (red triangles). {The averaged $\varepsilon_0(T)$ experimental dependence is shown by a thin red curve.} Calculated energy of the spin resonance $\omega_R$ (blue dots) is shown together with the uncertainty in its determination (inversely proportional to the spin resonance peak height, wide light blue region around the blue curve). \label{fig:GdResonance}}
\end{figure}

As the temperature increases, all gap features move towards the zero bias being directly associated with the $\Delta_{L,S}(T)$ temperature dependencies, which are shown in Figure~\ref{fig:GdResonance}. The large gap trend generally follows a single-gap BCS-like behavior except the notable curving down that starts at about 15\,K. Simultaneously, the small gap rapidly decreases and then expectedly steadily fades till the $T_c^\mathrm{local}$. Obviously, the different temperature trend indicates that the resolved energy parameters are related with two distinct superconducting condensates coexisting in Gd-1111. Additionally, this is the reason to interpret the dips at $\pm 3.2$\,mV as relating to the small gap rather than to the foot (otherwise, its temperature behavior would be similar to $\Delta_L(T)$~\cite{Gunsenheimer1994}). The observed temperature dependencies of the gap cannot be simulated in any conventional single-band model, however it is typical for the variety of the 1111 family of the FeBS studied before and could be described in the framework of the two-band model~\cite{Kuzmicheva2014,Kuzmicheva2017Sm} or the three-band model~\cite{Daghero2020}.

Beside the parent SGS, we resolved a fine structure caused by a resonant boson emission along with the IMARE process. Accompanying the SGS for the large gap, the less intensive dips appear at $|eV| \approx 33.4,~17.1$\,mV (red arrows in Figure \ref{temp}b). The fine structure is related to the resonant coupling with a characteristic bosonic mode. The boson energy harmonics $\varepsilon_0/n$ are therefore the `distances' between each satellite and the parent SGS dip, see the expression for $eV_n$ above, as illustrated by double arrows. In order to determine the boson energy, we extract it as an average distance to all $\Delta_L$ subharmonics, $\varepsilon_0 = \left\langle (eV_\mathrm{res_1} - eV_{n_L=1}) + 2 \cdot (eV_\mathrm{res_2} - eV_{n_L=2}) \right\rangle /4 \approx 10.4$\,meV, where $n_L$ and $n_\mathrm{res}$ are numbers of subharmonics corresponding to $\Delta_L$ and boson energy, respectively.

The satellites, however, are less pronounced as compared with the parent $\Delta_L$ dips, therefore, just a portion of carriers undergoing the multiple Andreev reflections emit a boson with energy $\varepsilon_0$. The position of the satellites agrees well between various SnS junctions, does not depend on the contact area and resistance, thus cannot be an artifact or caused by any dimensional effect. The observed fine structure {does} not match either $2\Delta_{L,S}/en$ nor $(\Delta_L + \Delta_S)/en$ subharmonic sequence, being reproducible from one sample to another, and independent {of} the contact dimension or any surface influence. This agrees well with the large statistics collected by us earlier with the break-junction probes of the oxypnictides of various {compositions}~\cite{Kuzmichev2017,Kuzmichev2017boson}.

In Figure~\ref{fig:GdResonance}, we show the temperature dependence of {superconducting} gaps {(violet and orange symbols)} together with the extracted boson energy $\varepsilon_0(T)$ {(red symbols; thin red line shows the averaged data)}. The values of the small gap were extrapolated for temperatures above 40\,K since they cannot be uniquely determined from the {experimental} spectra. $\varepsilon_0(T)$ follows neither $\Delta_{L,S}(T)$ nor the temperature dependence of their sum {(see green rhombuses in Figure~\ref{fig:GdResonance})}.
In addition, the specific temperature trend of {the fine structure position $eV_\mathrm{res}(T)$} or $\varepsilon_0(T)$ cannot simulate $\Delta(T)$ in the framework of any conventional model. For this reason, we also cannot attribute the satellites to a possible 
anisotropy of $\Delta_L$ in the momentum space.
Neither does the satellite structure relate with a distinct gap
, the largest order parameter (with the BCS ratio about 8), since the presence of three distinct gaps was not established for {the} 1111 oxypnictide family (for a review, see~\cite{JohnstonReview,Kuzmicheva2014,Kuzmicheva2017Sm,Si2016}).

Now that we proved that the origin of the observed anomaly is an intrinsic effect and {does} not directly {originate} from the gaps, we exclude two possible candidates for the role of the boson, namely, phonons and Leggett plasmons.

\textls[-10]{The value of $\varepsilon_0$ observed by us is close to the lowest-frequency optical phonon mode $\hbar\omega_\mathrm{phonon}$ = 11--14 meV unveiled in optimally doped 1111 compounds based on various lanthanides, with the highest $T_c \gtrsim 50$\,K~\cite{Zhao2008Raman,LeTacon2008,Christianson2008}. Obviously, the energy of this optical phonon mode must be dependent neither on the doping level (and thus on $T_c$ of the superconductor) nor on temperature within $T < T_\mathrm{Debye}$ \cite{Marini2008}. On the contrary, the experimentally observed bosonic mode energy $\varepsilon_0(T)$ weakly decreases until $T_c$ (see the red curve in Figure~\ref{fig:GdResonance}). At $T \to 0$, the energy $\varepsilon_0$ roughly scales with $T_c$ (see Figure~4 and Table~1 in \cite{Kuzmichev2017boson}) together with the superconducting gap values $\Delta_L(0)$ and $\Delta_S(0)$, as shown by us earlier for Gd and Sm-based oxypnictides with various doping levels (see Figure~12 in \cite{Kuzmicheva2017Sm} and Figure~5 in \cite{Kuzmicheva2014}). Therefore, we conclude the observed spectral feature has a non-phononic origin.}


Contrary to the case of MgB$_2$~\cite{Ponomarev2004,Ponomarev2007}, one cannot attribute the observed bosonic mode to the Leggett plasma mode~\cite{Leggett1966} mainly because several theoretical studies have shown that, {due to a moderate crossband interaction in the 1111-family compounds}, Leggett plasmons would have too large energy {(exceeding gap edge $2\Delta(0)$)} and therefore be unobservable in iron pnictides~\cite{Burnell2010,Ota2011}. 

\begin{figure}[H]
\includegraphics[width=0.8\linewidth]{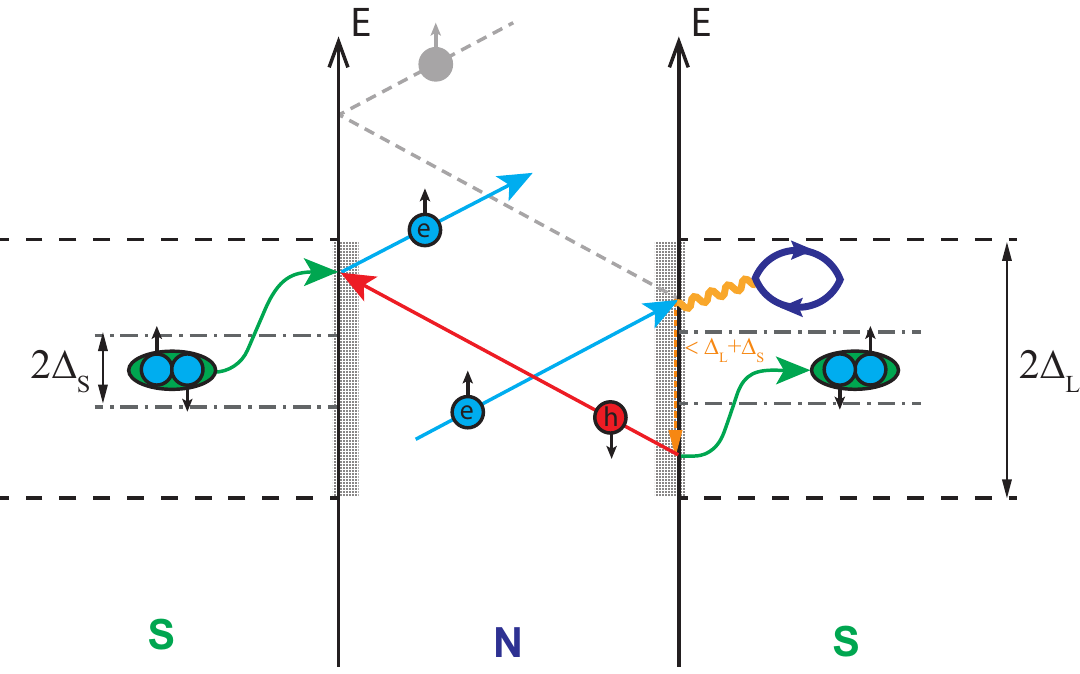}
\caption{Scheme of the multiple Andreev scattering where electron from the normal metal (N) within the proximity range near the superconductor (S) can scatter on the spin exciton (shown as the electron-hole bubble), lose energy $\varepsilon_0 < \Delta_L+\Delta_S$, and then form a Cooper pair inside the superconductor. Andreev reflected hole travels to the other side of the normal metal and, by annihilating with one electron from a Cooper pair of the left-hand side superconductor, results in the reflection of \linebreak the electron. \label{fig:ARfig}}
\end{figure}

\begin{figure}[H]
\includegraphics[width=0.8\linewidth]{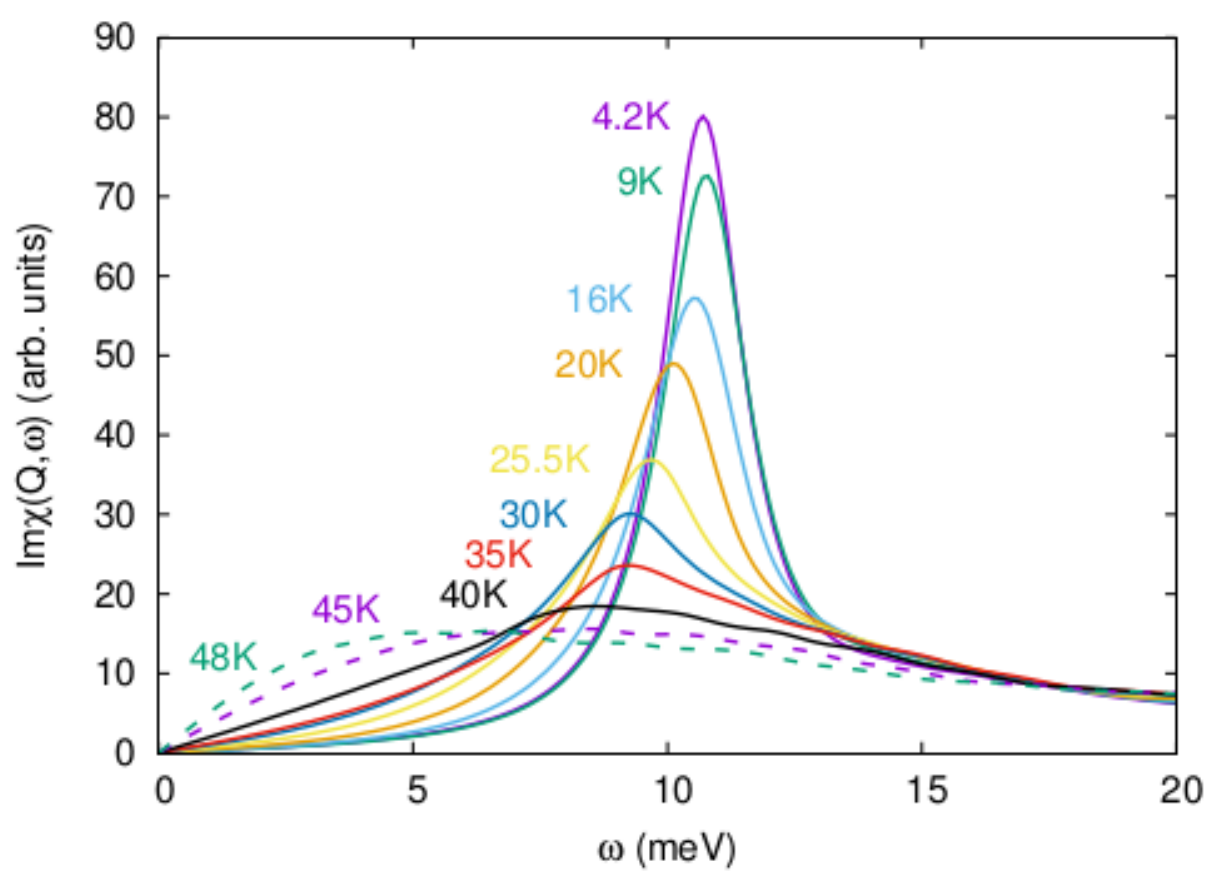}
\caption{Calculated temperature evolution of the imaginary part of the RPA spin susceptibility for the slightly anisotropic $s_\pm$ superconductor. Peak positions determine the spin exciton \linebreak frequency $\omega_R(T)$.
\label{fig:GdImChi}}
\end{figure}

The possibility left open is the scattering on the spin exciton that is formed in the superconductor. The process is sketched in Figure~\ref{fig:ARfig}. Within the random phase approximation (RPA), the energy of the spin exciton is calculated as a position of the spin resonance peak. Later it appears as a true divergence of the RPA spin susceptibility $\mathrm{Im}\chi(\mathbf{q},\omega)$ in the superconducting state with the $s_\pm$ gap~\cite{KorshunovEreminResonance2008,Maier2008}. {Iron-based materials are one of the interesting examples where RPA results for the pairing agree quite well with the more sophisticated theories. Later includes analytical (logarithmic) renormalization group (RG)~\cite{Chubukov2008,CvetkovicVDW2009,ChubukovReview,Maiti2010}, functional renormalization group (fRG)~\cite{r_thomale_09,Wang2010,Wang2011,Thomale2011,Classen2017}, and DFT+DMFT approach~\cite{Yin2014}. The origin of this agreement was extensively discussed earlier~\cite{MaitiKorshunovPRL2011,MaitiKorshunovPRB2011,Korshunov2014eng}. Taking into account that the electron--phonon interaction seems to be weak in pnictides~\cite{Boeri_08}, this leads to the conclusion that the RPA in application to the multiband Hubbard model provides a quite reasonable approach to the physics of iron-based materials~\cite{HirschfeldKorshunov2011}.} We have calculated the temperature dependence of the spin susceptibility within the five-orbital model for pnictides~\cite{Graser2009} using the measured values of the gaps as input parameters. Small gap anisotropy $\sim 10\%$ on the electron Fermi surface sheets was introduced similar to Ref.~\cite{Korshunov2018}. Depending on the set of on-site Coulomb interaction parameters, namely, Hubbard repulsion $U$, interorbital repulsion $U'$, Hund's $J$, and pair hopping $J'$, the peak slightly shifts in frequency and its width changes. The result for $U=1.4$\,eV, $U'=1$\,eV, and $J=J'=0.2$\,eV 
is shown in Figure~\ref{fig:GdImChi}. Positions of the maxima form the $\omega_R(T)$ dependence that is plotted in Figure~\ref{fig:GdResonance}.
Since, with the increasing temperature, the peaks become broader and their amplitudes diminish, the uncertainty in the spin exciton frequency increases.

\vspace{-9pt}

Another, indirect, contribution to the fine structure of the Andreev reflection from the spin exciton can come from the corresponding specific features of density of states (DOS). Those features, originating from the resonant coupling to the spin exciton at $T < T_c$, have the 
form of peaks at some energies and produce maxima of the Andreev reflection probability at the same energies (see~\cite{Octavio1983}, and Equations (8) and (9) in~\cite{Aminov1996}). As a result, the Andreev current increases at certain energies, thus in turn causing {the} appearance of the dynamic conductance features of the rather small amplitude ($< 0.08 G_N$) at roughly the same positions as the boson emission discussed above. Therefore, although the main contribution to the Andreev reflection comes from the coupling to the spin resonance mode, one should bear in mind some contribution of the indirect influence of the spin exciton through the DOS features.

\section{Conclusions}

In summary, a fine structure accompanying the {large superconducting gap}
$\Delta_L$ features is observed in dI(V)/dV spectra of clean classical (``long'') planar SnS-Andreev junctions in the GdO$_{0.88}$F$_{0.12}$FeAs superconductor of almost optimal composition with $T_c = 50$\,K. 
We have shown that this intrinsic effect is not directly related to the gaps themselves and excluded phonons and Leggett plasmons as the possible candidates for the role of the boson. Comparison to the theoretical calculations confirms that the observed feature originates from the scattering on the spin exciton that is formed in the superconducting state. Apart from that, the small contribution may come from the indirect effect of the spin resonance mode through the changes in the density of states. Spin resonance peak was seen earlier only in the inelastic neutron scattering. Thus, we provide an independent direct confirmation of the spin exciton appearance in Andreev spectra and prove the sign-changing $s_\pm$ gap structure in the 
studied FeBS.

\vspace{6pt}

\authorcontributions{Conceptualization, M.M.K. and S.A.K.; calculations, M.M.K.; experiment, T.E.K. and S.A.K.; writing, M.M.K., T.E.K., and S.A.K.; funding acquisition, M.M.K., T.E.K., and S.A.K. All authors have read and agreed to the published version of the manuscript.}

\funding{All authors acknowledge support by the state assignment of the Ministry of Science and Higher Education of the Russian Federation, nos. 0023-2019-0005 and 0287-2021-0035.
}

\institutionalreview{Not applicable.}

\informedconsent{Not applicable.}

\dataavailability{The datasets generated during and/or analysed during the current study are available from the corresponding author on reasonable request.}

\acknowledgments{We acknowledge useful discussions with A.S. Melnikov, S.G. Ovchinnikov, and V.M. Pudalov. {S.A.K. and T.E.K. are grateful to the late E.P. Khlybov for the provided Gd-1111 samples.} The research has been partly conducted using the research equipment of the shared facility center at P.N. Lebedev Physical Institute RAS.}

\conflictsofinterest{The authors declare no conflict of interest.}



\begin{adjustwidth}{-\extralength}{0cm}

\reftitle{References}



\end{adjustwidth}

\end{document}